\documentclass[a4paper,12pt]{article}
\usepackage[T1]{fontenc}
\usepackage[utf8]{inputenc}
\pdfoutput=1
\usepackage{amsmath}
\usepackage{amsfonts}
\usepackage{amssymb}
\usepackage[english]{babel}
\usepackage[pdftex]{graphicx,color}
\usepackage[mathscr]{eucal}

\begin{document}
\title{BPS center vortices in nonrelativistic $SU(N)$ gauge models with adjoint Higgs fields}
\author{~L.~E.~Oxman
\\ \\ Instituto de F\'{\i}sica, Universidade Federal Fluminense,\\
Campus da Praia Vermelha, Niter\'oi, 24210-340, RJ, Brazil.}
\date{\today}
\maketitle 
 
\begin{abstract} 
In this work, we propose a class of $SU(N)$ Yang-Mills models, with 
adjoint Higgs fields, that accept BPS center vortex equations. 
The lack of a \textit{local} magnetic flux that
could serve as an energy bound is circumvented by including a new
term in the energy functional. This term tends to align, in the Lie
algebra, the magnetic field and one of the adjoint Higgs fields. Finally,
a reduced set of equations for the center vortex profile functions
is obtained (for $N=2,3$). In particular, $Z(3)$ BPS vortices come
in three colours and three anticolours, obtained from an ansatz based
on the defining representation and its conjugate.
\end{abstract} 
\textbf{Keywords}: \\

\textbf{Pacs}: 11.15.-q, 11.10.Lm, 11.15.Kc

\section{Introduction} 

Topological solitons are present in many areas of Physics. Some well-known
examples are: kinks in polyacetylene \cite{Sch}, vortices in type
II superconductors, skyrmions in magnetic systems \cite{Fradkin},
and skyrmions to describe baryons in flavour symmetric models \cite{skyrme,skyrme-rational}.
To gain information about these objects, it is important to identify
a critical point where a BPS bound is obtained.
Namely, a point where the energy can be written as a sum of squares
plus the topological charge of the field configuration. As continuous
field deformations cannot modify this charge, setting the squares
to zero leads to a set of (BPS) equations whose solutions are absolute
minima in the given topological sector. In this process, the equations
are reduced to first order, which facilitates analytical and numerical
studies of these systems. In addition, BPS multisoliton solutions
with a given total charge have the same energy, so the forces between
BPS solitons vanish. For these reasons, the critical point provides
a nice reference to introduce perturbations and study the soliton
dynamics \cite{Manton}.

Topological solitons are also important in effective descriptions
of the strong interactions. Abelian Higgs models have been proposed
to describe the $q\bar{q}$ potential \cite{Baker,Baker1} and the
interaction among three quarks \cite{MSu}-\cite{C1}. In refs. \cite{deVega}-\cite{HV},
center vortices were accommodated in $SU(N)$ Yang-Mills models with
$N$ adjoint Higgs fields; these objects can describe the N-ality
properties of the confining string \cite{Konishi-Spanu}. Recently,
we proposed a class of flavour symmetric models supporting not only the confining string between a $q\bar{q}$ colourless pair of external quarks, but also
other possible excited states \cite{conf-qg}. Among them, $qg\bar{q}'$
hybrid mesons \cite{hybrids}-\cite{DuE}, formed for example by a red/anti-green
pair of quarks bound by an anti-red/green valence gluon. While the
normal string is a center vortex of the effective model, the excited
string is formed by a pair of center vortices interpolated by a monopole,
which is identified with a confined valence gluon. 

The topology and classification of center vortices have been analyzed
in ref. \cite{Konishi-Spanu}, when a general compact gauge group $G$ is broken down to its center. The roots of the
Lie algebra and the weights of their representations play an important
role, as occurs when characterizing non Abelian monopoles \cite{GNO}.
BPS equations for non Abelian vortices have been obtained in refs.
\cite{David1}-\cite{David}, for a review, see refs. \cite{vortices-d,notes}.

In ref. \cite{conf-qg}, we proposed a Lorentz invariant flavour symmetric model that is expected to contain center vortices, as it possesses the proper SSB pattern and topology. The problem is that the field equations are mathematically difficult. They can only be solved by following numerical methods. With the aim of exploring the usual tools to understand topological objects, in this work we shall look for models accepting BPS center vortices, governed by {\it first order} field equations. As an intermediate step, we shall simplify the content of the flavour symmetric 
model, which is based on $N^{2}-1$ adjoint Higgs fields that form a local Lie basis at the nontrivial
vacua of the Higgs potential. Observing that the
essential features of the Lie algebra can be captured by a reduced
set of fields and conditions, labelled by the simple roots, a Lorentz invariant model that for $N\geq 3$ has a simplified field content will be obtained. The Higgs potential will be such that its minimization returns a set of conditions that essentially define a Chevalley basis. This model has the same SSB pattern and topology than the former. Next, we shall make an extension to obtain a model that accepts BPS center vortices. 
In this process, we can anticipate some peculiarities. Generally, BPS equations are 
derived by working on the energy functional to obtain a bound (for
an alternative approach, see ref. \cite{B-GP}). For $U(1)$ vortices,
the bound is given by the magnetic flux. This is a topological term
that can be written locally, by means of a flux density, so it can
indeed arise by working on the energy, which is a local functional.
On the other hand, for center vortices, the flux concept is replaced
by the asymptotic behavior of the gauge invariant Wilson loop, a nonlocal
object that may not appear in the energy calculation. Then, the search
for BPS equations in $SU(N)\to Z(N)$ SSB models led us to consider
the introduction of a nonrelativistic interaction term that tends to align, in the
Lie algebra, the magnetic field along one of the adjoint Higgs
fields. This in turn implied a different type of bound. After completing
the squares, the energy is always greater than or equal to zero. Thus,
BPS center vortices are nonrelativistic objects characterized by an exact compensation between
the positive definite part of the energy functional (kinetic energy
plus Higgs potential) and the Lie algebra alignment contribution.

In this regard, two comments are in order:  i) The reason for considering the intermediate step is that, for $N \geq 3$, the direct inclusion of the alignment term in the flavour symmetric model would lead, after completing the squares, to too many (possibly incompatible) conditions to saturate the bound. ii) 
Because of Lorentz symmetry breaking, and rotational symmetry breaking in $3+1$ dimensions, the BPS models are not directly physically relevant. However, the presence 
of a BPS point in the extended parameter space could serve as a check for the numerical analysis, when moving away from the physically relevant non BPS Lorentz invariant confining models.

The general BPS solution will be written in terms of a set of profile
functions and a mapping $R(S)$ in the adjoint representation of $SU(N)$.
The mapping $S\in SU(N)$ contains information about the asymptotic
Wilson loop and the possible defects at the vortex guiding centers,
which determine the profile behaviours. Because of the model's topology,
a given phase $S_{0}(\varphi)$, defined close to and around a vortex
guiding center, can be extended to different asymptotic phases $S_{a}(\varphi)$,
where $S_{a}(\varphi+2\pi)=e^{i2\pi z_{a}/N}S_{a}(\varphi)$, $z_{a}\in\mathbb{Z}$.
The $Z(N)$ charge is due to the fact that the different extensions
are related by $(z_{a'}-z_{a})/N\in\mathbb{Z}$. For the same reason,
a given $S_{a}$ can be matched with different phases $S_{0}$, with their
respective pointlike defects.
When leaving the critical point, by lowering the alignment interaction
term, some of these extensions will become unstable. For example,
for vanishing $Z(N)$-charge the defect can be avoided, and the lowest
energy solution will simply correspond to a trivial regular gauge
transformation of the SSB vacua. For $Z(N)$ charge $\pm1$, we shall
discuss the BPS solutions that are expected to be related to the stable
$Z(2)$ and $Z(3)$ noncritical center vortices.

The article is organized as follows. In section 2, we construct the
simplified $SU(N)$ model and discuss the possible vacua. In section 3, we obtain
the bounds and the set of BPS equations (for $N=2,3$). Some properties
of the field parametrization are discussed in section 4. Section 5
is devoted to obtaining information about the BPS solutions and 
discussing the BPS center vortex. Finally, in section 6, we present our conclusions.

\section{Models with $SU(N)\to Z(N)$ SSB}

In order to support center vortices, we are interested in driving a phase where the gauge symmetry is spontaneously broken down to $Z(N)$. For example,  in ref. \cite{conf-qg}, we introduced a model displaying a flavour symmetry.
That is, we considered the energy functional\footnote{We are using the inner product $\langle X,Y\rangle=Tr\left(Ad(X)^{\dagger}Ad(Y)\right)$, where $Ad(\cdot)$ is a linear map into the adjoint representation.}, 
\begin{eqnarray}
E & = & \int d^{3}x\,\left(\frac{1}{2}\langle B_{i}\rangle^{2}+\frac{1}{2}\langle D_{i}\psi_{A}\rangle^{2}+V_{{\rm Higgs}}(\psi_{A})\right)\;,\label{mflavour}
\end{eqnarray}
where $B_i$ is the chromomagnetic field, 
$D_i=\partial_i-ig[A_i,~]$, and the potential for the hermitian adjoint Higgs fields $\psi_{A}$, $A=1,\dots,d=N^2-1$, is given by, 
\begin{eqnarray}
V_{{\rm Higgs}} & = & c+\frac{\mu^{2}}{2}\,\langle\psi_{A},\psi_{A}\rangle + \frac{\kappa}{3}\, f_{ABC}\langle (-i)[\psi_{A}, \psi_{B}],\psi_{C}\rangle+\frac{\lambda}{4}\,\langle [\psi_{A},\psi_{B}],[\psi_{A},\psi_{B}]\rangle\;,\nonumber \\
\label{Vf}
\end{eqnarray}
where $f_{ABC}$ are structure constants of the $\mathfrak{su}(N)$ Lie algebra.
At $\mu^{2}=\frac{2}{9}\frac{\kappa^{2}}{\lambda}$,
$\kappa<0$, we can write, 
\begin{eqnarray}
V_{{\rm Higgs}} & = & \frac{\lambda}{4}\,\langle\Psi_{AB}\rangle^{2}\makebox[.5in]{,}\Psi_{AB}=f_{ABC}\, v_{c}\,\psi_{C}+i[\psi_{A},\psi_{B}]\;,
\end{eqnarray}
\begin{equation}
v_{c}=-\frac{\kappa}{2\lambda}\pm\sqrt{\left(\frac{\kappa}{2\lambda}\right)^{2}-\frac{\mu^{2}}{\lambda}}=-\frac{2\kappa}{3\lambda}\;,
\end{equation}
after adjusting $c$, so that the potential energy for vacuum configurations
vanishes. The space of vacua ${\cal M}$ is obtained from the conditions
$\Psi_{AB}=0$, i.e., 
\begin{equation}
[\psi_{A},\psi_{B}]=if_{ABC}\, v_{c}\,\psi_{C}\;.
\end{equation}
This encompasses the trivial symmetric point $\psi_{A}=0$, separated
by a potential barrier from the nontrivial points. Of course, starting
from a nontrivial point $\psi_{A}\in{\cal M}$, we can generate a
continuum $S\psi_{A}S^{-1}$, $S\in SU(N)$, that is also in ${\cal M}$.
In addition, the only transformations that leave these points invariant
are $S\in Z(N)$, so they correspond to $SU(N)\to Z(N)$ SSB vacua.
For $N\geq3$, the SSB points can be divided into a pair of distinct
sets, separated by a potential barrier, corresponding to the defining
representation and its conjugate, 
\begin{equation}
\psi_{A}=ST_{A}S^{-1}\makebox[.5in]{,}\psi_{A}=S(-T_{A})^{\ast}S^{-1}\;.
\end{equation}
For $N=2$, this pair collapses into a single component, as a matrix 
$S_{c}\in SU(2)$ exists such that $(-T_{A})^{\ast}=S_{c}T_{A}S_{c}^{-1}$,
$A=1,2,3$.

Although the model in eq. (\ref{mflavour}) contains center vortices, we did not succeed in taking it as a starting point to obtain BPS equations (for $N\geq 3$). The presence of too many fields ultimately leads to incompatible conditions to saturate the bound. For this reason, in the next section, we shall look for a simplified model. Instead of the previous $N^2-1$ hermitian fields, we  introduce $N-1$ hermitian and $N-1$ complex adjoint Higgs fields. For 
$N \geq 3$, this will result in a simpler set of fields and conditions to define the SSB vacua. This, together with the ``alignment'' term introduced in section \ref{BPSmodels}, will finally lead to a set of compatible BPS equations.   

\subsection{Simplified model}

Let us consider hermitian variables, $\psi_{q}$, $q=1,\dots,r=N-1$,
and complex variables $\zeta_{\alpha}$, labelled by the positive
simple roots $\vec{\alpha}_{q}$ ($\vec{\alpha}_{1}<\vec{\alpha}_{2}\dots<\vec{\alpha}_{r}$).
The conditions, 
\begin{equation}
[\psi_{q},\psi_{p}]=0\makebox[.5in]{,}v_{c}\,\vec{\alpha}|_{q}\,\zeta_{\alpha}-[\psi_{q},\zeta_{\alpha}]=0\;,\label{cond1}
\end{equation}
contain most of the relevant structure of the Lie algebra. For nontrivial
fields $\zeta_{\alpha}$, we can imply,
\begin{itemize}
\item i) The fields $\psi_{q}$ are nontrivial, as their sizes are fixed by the
eigenvalues $v_{c}\,\vec{\alpha}|_{q}$. They are also linearly independent: if there is a combination
$\gamma^{q}\,\psi_{q}=0$, then using eq. (\ref{cond1}) we get $\vec{\gamma}\cdot\vec{\alpha}=0$,
for every simple root, so that $\vec{\gamma}=0$.
\item ii) $[\zeta_{\alpha},\zeta_{\alpha}^{\dagger}]$ is in the Cartan
subalgebra generated by the fields $\psi_{q}$.
\item iii) As the positive (negative) roots can be written as a linear combination
of the simple roots, with nonnegative (nonpositive) integer coefficients, any root vector is proportional to an appropriate chain of
operations of the form $[\zeta_{\alpha},\zeta_{\alpha'}]$ ($[\zeta_{\alpha}^{\dagger},\zeta_{\alpha'}^{\dagger}]$).
\item iv) As the difference of a pair of positive simple roots cannot be
a root, we have $[\zeta_{\alpha},\zeta_{\alpha'}^{\dagger}]=0$.
\end{itemize}
However, considering a potential whose minimization only leads to
the conditions in eq. (\ref{cond1}) would not be the desired one.
Given a nontrivial solution $(\psi_{q}$, $\zeta_{\alpha})$, the
replacement $\zeta_{\alpha}\to t\,\zeta_{\alpha}\,$, $t\in\mathbb{R}$,
would also lead to a solution. Then, the interesting SSB initial point
could be continuously moved to $(\psi_{q},0)$, and then $\psi_{q}$
could be continuously moved to $0$, always staying in the space of
vacua ${\cal M}$. That is, there would be no potential barrier between
the interesting configurations and the trivial one. This will be corrected
by including a term in the potential to avoid, after minimization,
the possibility of moving the fields $\zeta$ to zero, when we start with
a SSB point. 
For this purpose, let us consider the additional condition, 
\begin{equation}
\sum_{q}s_{q}\,\left(v_{c}\,\vec{\alpha}_{q}\cdot\vec{\psi}-[\zeta_{\alpha_{q}},\zeta_{\alpha_{q}}^{\dagger}]\right)=0\makebox[.5in]{,}\vec{\alpha}\cdot\vec{\psi}=\vec{\alpha}|_{q'}\psi_{q'}\;,\label{addc}
\end{equation}
where $s_{q}$ takes values $+1$ or $-1$. Now, we consider a solution
to eq. (\ref{cond1}), and recall that given linearly independent
fields $\psi_{q}$ it is always possible to introduce unique elements $\mathscr{H}_{\alpha}$
such that, 
\begin{equation}
\langle\mathscr{H}_{\alpha},\psi_{q}\rangle=v_{c}\,\vec{\alpha}|_{q}\;.\label{varia}
\end{equation}
As is well known \cite{Humphreys}-\cite{Giorgi}, these variables satisfy, 
\begin{equation}
[\zeta_{\alpha},\zeta_{\alpha}^{\dagger}]=\langle\zeta_{\alpha},\zeta_{\alpha}\rangle\,\mathscr{H}_{\alpha}\;.\label{Hal1}
\end{equation}
Using this information in eq. (\ref{addc}) and projecting
with $\mathscr{H}_{\alpha_{p}}$, we get, 
\begin{eqnarray}
\lefteqn{\sum_{q}s_{q}\left(v_{c}\,\vec{\alpha}_{q}|_{q'}\,\langle\mathscr{H}_{\alpha_{p}},\psi_{q'}\rangle-\langle\zeta_{\alpha_{q}},\zeta_{\alpha_{q}}\rangle\,\langle\mathscr{H}_{\alpha_{p}},\mathscr{H}_{\alpha_{q}}\rangle\right)=}\nonumber \\
 &  & =\sum_{q}s_{q}\left(v_{c}^{2}\,\vec{\alpha}_{q}\cdot\vec{\alpha}_{p}-\langle\zeta_{\alpha_{q}},\zeta_{\alpha_{q}}\rangle\,\langle\mathscr{H}_{\alpha_{p}},\mathscr{H}_{\alpha_{q}}\rangle\right)=0\;.\label{term1}
\end{eqnarray}

From the Lie algebra internal product and the mapping $\vec{\alpha}\to\mathscr{H}_{\alpha}$,
an internal product on the root space can be defined \cite{Humphreys}-\cite{Giorgi},
\begin{equation}
\langle\vec{\alpha},\vec{\alpha}\,'\rangle\equiv\langle\mathscr{H}_{\alpha},\mathscr{H}_{\alpha'}\rangle\;.
\end{equation}
These quantities are strongly constrained. In particular, 
\begin{equation}
2\,\langle\vec{\alpha},\vec{\alpha}\,'\rangle/\langle\vec{\alpha}\,',\vec{\alpha}\,'\rangle\in\mathbb{Z}\;,\label{Ci}
\end{equation}
are the so-called Cartan integers, which determine the geometry of
the root lattice. They do not depend on the Cartan basis, coinciding
with, 
\begin{equation}
2\,(\vec{\alpha}\cdot\vec{\alpha}\,')/(\vec{\alpha}\,'\cdot\vec{\alpha}\,')\;,
\end{equation}
which corresponds to (\ref{Ci}), when computed with an orthogonal basis $\psi_{q}$,
$\langle\psi_{q},\psi_{p}\rangle=v_{c}^{2}\,\delta_{qp}$. Note that
in this case, $\mathscr{H}_{\alpha}=(1/v_{c})\,\vec{\alpha}|_{q}\psi_{q}$.
In addition, for $\mathfrak{su}(N)$, the lengths of the roots are equal, $\langle\vec{\alpha},\vec{\alpha}\rangle=c$,
$\vec{\alpha}\cdot\vec{\alpha}=1/N$. Then, using this information,
eq. (\ref{term1}) implies, 
\begin{equation}
\sum_{q}s_{q}\,\left(v_{c}^{2}\,\vec{\alpha}_{q}\cdot\vec{\alpha}_{p}-Nc\,\langle\zeta_{\alpha_{q}},\zeta_{\alpha_{q}}\rangle\,\vec{\alpha}_{q}\cdot\vec{\alpha}_{p}\right)=0\;.
\end{equation}
This is valid for any basis element $\vec{\alpha}_{p}$, that is,
\begin{equation}
\sum_{q}s_{q}\,\left(v_{c}^{2}-Nc\,\langle\zeta_{\alpha_{q}},\zeta_{\alpha_{q}}\rangle\right)\,\vec{\alpha}_{q}=0\;,
\end{equation}
and as the simple roots $\vec{\alpha}_{q}$ are linearly independent, we get, 
\begin{equation}
\langle\zeta_{\alpha_{q}},\zeta_{\alpha_{q}}\rangle=\frac{v_{c}^{2}}{Nc}\;.\label{prot}
\end{equation}
This means that if we define the space of vacua ${\cal M}$ by means
of the conditions (\ref{cond1}) and (\ref{addc}), a nontrivial SSB point $(\psi_{q},\zeta_{\alpha})\in{\cal M}$ cannot be continuously moved to the trivial solution $(0,0)$, always staying in ${\cal M}$. In effect, starting with nontrivial fields
$\zeta_{\alpha}$ implies linearly independent fields $\psi_{q}$, and
this in turn leads to nontrivial $\mathscr{H}_{\alpha}$ that protect
the size of $\zeta_{\alpha}$ through eq. (\ref{prot}). In other
words, a potential whose minimization gives the conditions (\ref{cond1})
and (\ref{addc}) has a barrier between the SSB vacua and the trivial
one.

It is convenient to introduce a model without referring to a particular
convention for the simple roots. The field $\vec{\psi}$ with $r=N-1$
components, such that $\vec{\psi}|_{q}=\psi_{q}$, can be expanded
either in terms of the simple roots $\vec{\alpha}_{q}$ or the $\vec{\lambda}^{q}$
basis satisfying, 
\begin{equation}
\vec{\alpha}_{q}\cdot\vec{\lambda}^{p}=\delta_{q}^{\, p}\makebox[.5in]{,}\vec{\alpha}_{q}|_{p}\,\vec{\lambda}^{q}|^{p'}=\delta_{p}^{\, p'}\;,\label{alam}
\end{equation}
\begin{equation}
\vec{\psi}=\vec{\alpha}_{q}\,\phi^{q}\makebox[.8in]{{\rm with,}}\phi^{q}=\vec{\lambda}^{q}\cdot\vec{\psi}\;,
\end{equation}
\begin{equation}
\vec{\psi}=\vec{\lambda}^{q}\,\phi_{q}\makebox[.8in]{{\rm with,}}\phi_{q}=\vec{\alpha}_{q}\cdot\vec{\psi}\;,
\end{equation}
\begin{equation}
\psi_{p}=\vec{\alpha}_{q}|_{p}\,\phi^{q}=\vec{\lambda}^{q}|_{p}\,\phi_{q}\;.
\end{equation}
For $\mathfrak{su}(N)$, the $\vec{\lambda}^{q}$ basis is given by,
\begin{equation}
\vec{\lambda}^{q}=2N\,\vec{\Lambda}^{q}\;,
\end{equation}
where $\vec{\Lambda}^{q}$ are the fundamental weights (see appendix
A).

The relation between the different components is, 
\begin{equation}
\phi_{q}=A_{qp}\,\phi^{p}\makebox[.8in]{,}\phi^{q}=A^{qp}\,\phi_{p}\;,
\end{equation}
\begin{equation}
A_{qp}=\vec{\alpha}_{q}\cdot\vec{\alpha}_{p}\makebox[.5in]{,}A^{qp}=\vec{\lambda}^{q}\cdot\vec{\lambda}^{p}\;.
\end{equation}
For $\mathfrak{su}(N)$, the quantities $C_{qp}=2NA_{qp}$ are the
elements of the Cartan matrix, which define the natural product in
the root space.

With these definitions, together with $\zeta_{q}=\zeta_{\alpha_{q}}$,
the conditions in eqs. (\ref{cond1}) and (\ref{addc}) become, 
\begin{equation}
[\phi_{q},\phi_{p}]=0\makebox[.5in]{,}v_{c}\,\delta_{(p)}^{q}\,\zeta_{(p)}-[\phi^{q},\zeta_{p}]=0\;,\label{conda}
\end{equation}
\begin{equation}
\sum_{q}s_{q}\,\left(v_{c}\,\phi_{q}-[\zeta_{(q)},\zeta_{(q)}^{\dagger}]\right)=0\;,
\end{equation}
which can be obtained by minimizing the Higgs potential, 
\begin{equation}
V_{{\rm Higgs}}=\frac{1}{2}\langle\Phi,\Phi\rangle+\langle Z_{q},Z_{q}\rangle+{\cal R}\;,\label{potsim}
\end{equation}
where, 
\begin{equation}
\Phi=\sum_{q}s_{q}\Phi_{q}\makebox[.5in]{,}\Phi_{q}=\sqrt{\gamma}\left(v_{c}\,\phi_{q}-[\zeta_{(q)},\zeta_{(q)}^{\dagger}]\right)\;,
\end{equation}
\begin{equation}
Z_{q}=\sqrt{\gamma_{z}}\left(v_{c}\,\zeta_{q}-[\phi^{(q)},\zeta_{(q)}]\right)\;,
\end{equation}
\begin{equation}
{\cal R}=\gamma_{r}\sum_{q\neq p}\left(\frac{1}{2}\langle[\phi_{q},\phi_{p}]\rangle^{2}+\langle[\phi^{q},\zeta_{p}]\rangle^{2}\right)\;.
\end{equation}
Now, noting that, 
\begin{equation}
\langle D_{i}\psi_{q},D_{i}\psi_{q}\rangle=A^{qp}\langle D_{i}\phi_{q},D_{i}\phi_{p}\rangle\;,
\end{equation}
we initially propose the model, 
\begin{equation}
E=\int d^{3}x\,\left(\frac{1}{2}\langle B_{i}\rangle^{2}+\frac{1}{2}\langle D_{i}\phi_{q},D_{i}\phi^{q}\rangle+\langle D_{i}\zeta_{q},D_{i}\zeta_{q}\rangle+V_{{\rm Higgs}}\right)\;.\label{Emod}
\end{equation}
Here, the space of vacua ${\cal M}$ is given by the trivial point
$\phi_{q}=0$, $\zeta_{q}=0$, separated by a potential barrier from
the SSB points. For $N\geq3$, the latter can be separated into the
sets, 
\begin{equation}
\phi_{q}=S(\vec{\alpha}_{q}\cdot\vec{H})S^{-1}\makebox[.5in]{,}\zeta_{q}=SE_{\alpha_{q}}S^{-1}\label{defi}
\end{equation}
\begin{equation}
\phi_{q}=S(-\vec{\alpha}_{q}\cdot\vec{H})S^{-1}\makebox[.5in]{,}\zeta_{q}=S(-E_{\alpha_{q}}^{T})S^{-1}\;,\label{conju}
\end{equation}
where $E_{\alpha}^{T}$ is the transpose of $E_{\alpha}$, and $H_{q}$,
$E_{\alpha}$ are elements of a Cartan basis (see appendix A). For
$N=2$, the sets in (\ref{defi}) and (\ref{conju}) are equal.

\section{Nonrelativistic models with BPS center vortex equations}
\label{BPSmodels}

As is well known, center vortices are characterized by a center element
\begin{equation}
\mathfrak{z}=e^{i2\pi z/N}I\in SU(N)
\end{equation}
such that, for a path linking the vortex and contained in the asymptotic
region, the Wilson loop gives, 
\begin{equation}
W[A]=\mathfrak{z}\;.\label{Wzeta}
\end{equation}
The center vortex has a $Z(N)$ charge given by $z$, defined modulo
$N$. In particular, this is the case when in an asymptotic region
$r>r_{m}$ the gauge field is given by, 
\begin{equation}
A_{i}=\frac{1}{g}\partial_{i}\varphi\,\vec{\beta}\cdot\vec{H}\makebox[.5in]{,}e^{i2\pi\,\vec{\beta}\cdot\vec{H}}=\mathfrak{z}\;,\label{poss-b}
\end{equation}
where $r$ and $\varphi$ are polar coordinates with respect to the
vortex axis. The possible magnetic weights $\vec{\beta}$ satisfy
\begin{equation}
\vec{\beta}\cdot\vec{\alpha}\in Z\;,\label{qc}
\end{equation}
for every root $\vec{\alpha}$. The solutions to eq. (\ref{qc}) are \cite{GNO}, 
\cite{Fidel2,Konishi-Spanu}, 
\begin{equation}
\vec{\beta}=2N\vec{w}\;,\label{solution}
\end{equation}
where $\vec{w}$ are the weights of the different representations.
The minimum charge center vortices ($z=\pm1$) can be labelled by
the weights of the defining representation and its conjugate \cite{Konishi-Spanu}.

In the asymptotic region, if a Higgs field
takes the value $X_{0}$, $\langle X_{0},X_{0}\rangle=v_{c}^{2}$
at $\varphi=0$ then, on the circle at infinity, the non Abelian phase
will accompany the $A_{i}$ behaviour in eq. (\ref{poss-b}) as follows\footnote{we use $X$ to denote any of the Higgs fields $\phi_q$, $\zeta_q$.}, 
\begin{equation}
X=SX_{0}S^{-1}\makebox[.5in]{,}S=e^{i\varphi\,\vec{\beta}\cdot\vec{T}}\;.\label{asyr}
\end{equation}

Now, we would like to propose models accepting BPS center vortex equations
for $SU(2)$ and $SU(3)$. To simplify the discussion, let us consider
planar systems, replacing $d^{3}x\to d^{2}x=dx_{1}dx_{2}$, $B_{3}\to B$,
and taking $B_{1}=B_{2}=0$.
Initially, we note that the type of models we have discussed so far
cannot accept a BPS bound. Indeed, this would be the case in any model
whose energy functional only vanishes for vacuum configurations, while
on the space of field configurations $\{A,X\}_{\mathfrak{z}}$,
with a given nontrivial asymptotic behaviour labelled by $\mathfrak{z}$,
it is strictly positive. In this case, to obtain BPS center vortex
equations, the energy functional should be bounded by a nonzero
term verifying: i) gauge invariance, ii) it assumes a fixed value
on the space$\{A,X\}_{\mathfrak{z}}$, that only depends on $\mathfrak{z}$
(topological), iii) as the bound would be derived by working on the
energy density, it should have the form $\int d^{2}x\,\rho$ (locallity).
While the Wilson loop verifies i) and ii), it is a nonlocal object
that cannot arise in the calculation. On the other hand, while 
\begin{equation}
\int d^{2}x\,\langle\eta,B\rangle\makebox[.5in]{,}B=\partial_{1}A_{2}-\partial_{2}A_{1}-ig[A_{1},A_{2}]\;,\label{termc}
\end{equation}
with $\eta$ an adjoint field, satisfies i) and iii), it does not satisfy ii). This would be a boundary term for homogeneous
$\eta$ and those Abelian-like fields in $\{A,X\}_{\mathfrak{z}}$
such that $B=\partial_{1}A_{2}-\partial_{2}A_{1}$ on the \textit{whole}
plane. Then, the search for BPS center vortices should consider a
modified class of models where configurations in $\{A,X\}_{\mathfrak{z}}$
do not necessarily have strictly positive energy.

For example, we will see that the model (\ref{Emod}) could be reorganized
as a sum of squares plus a term of the form (\ref{termc}). As this
term does not satisfy ii), setting the squares to zero will not produce,
in a given sector $\{A,X\}_{\mathfrak{z}}$, solutions to the field
equations associated with (\ref{Emod}). Then, it is natural to try a modified model, 
\begin{equation}
E=\int d^{2}x\,\left(\frac{1}{2}\langle B\rangle^{2}+\frac{1}{2}\langle D_{i}\phi_{q},D_{i}\phi^{q}\rangle+\langle D_{i}\zeta_{q},D_{i}\zeta_{q}\rangle+V_{{\rm Higgs}}-\langle\eta,B\rangle\right)\;.\label{energy}
\end{equation}
Here, we have included the gauge invariant $\langle\eta,B\rangle$-interaction,
that tends to align $B$ along $\eta$ in the Lie algebra. The field
$\eta$ will be an appropriate combination of the adjoint Higgs fields,
to be determined in order for the model to accept BPS center vortex
equations. At the critical point, we shall see that in spite of the
last term in eq. (\ref{energy}), this energy functional satisfies
$E\geq0$. For BPS solutions, the contribution originated from the
positive definite terms will be exactly compensated by the energy
lowering due to the Lie algebra alignment between magnetic and Higgs
fields. Thus, the topologically nontrivial BPS center vortices will
have vanishing energy. We note that with this term the planar model
becomes nonrelativistic although it continues to be isotropic in $2+1$ dimensions\footnote{In $3+1$ dimensions, this type of model would also break rotation symmetry.}.

Let us derive the fundamental property to discuss BPS bounds. Using the ciclicity of the internal product, 
\begin{equation}
\langle X,[Y,Z]\rangle=\langle[X,Z^{\dagger}],Y\rangle\;,\label{mix1}
\end{equation}
as $A_{i}$ is hermitian, we have, 
\begin{eqnarray}
\langle D_{i}X,Y\rangle & = & \langle\partial_{i}X-ig[A_{i},X],Y\rangle\nonumber \\
 & = & \partial_{i}\langle X,Y\rangle-\langle X,\partial_{i}Y\rangle+ig\langle[A_{i},X],Y\rangle\nonumber \\
 & = & \partial_{i}\langle X,Y\rangle-\langle X,D_{i}Y\rangle\;,\label{porpart}
\end{eqnarray}
\begin{eqnarray}
\langle D_{i}X,Y\rangle+\langle X,D_{i}Y\rangle & = & \partial_{i}\langle X,Y\rangle\;.\label{psipsi}
\end{eqnarray}
Now, defining, 
\begin{equation}
D=D_{1}+iD_{2}\;,
\end{equation}
we note that, 
\begin{eqnarray}
 &  & \langle DX,DX\rangle=\langle D_{1}X+iD_{2}X,D_{1}X+iD_{2}X\rangle\nonumber \\
 &  & =\langle D_{1}X,D_{1}X\rangle+\langle D_{2}X,D_{2}X\rangle-i\langle D_{2}X,D_{1}X\rangle+i\langle D_{1}X,D_{2}X\rangle\;.
\end{eqnarray}
In addition, as $B$ is hermitian, 
\begin{equation}
\langle X,[B,X]\rangle=\langle[X,X^{\dagger}],B\rangle=\langle B,[X,X^{\dagger}]\rangle\;.
\end{equation}
This together with eq. (\ref{porpart}) and, 
\begin{equation}
\left[D_{\mu},D_{\nu}\right]X=-ig[F_{\mu\nu},X]\;,
\end{equation}
which is obtained from the Jacobi identity, we get, 
\begin{eqnarray}
\lefteqn{\langle D_{2}X,D_{1}X\rangle-\langle D_{1}X,D_{2}X\rangle=}\nonumber \\
 &  & =-\langle X,D_{2}D_{1}X\rangle+\langle X,D_{1}D_{2}X\rangle+\partial_{2}\langle X,D_{1}X\rangle-\partial_{1}\langle X,D_{2}X\rangle\nonumber \\
 &  & =-ig\langle X,[F_{12},X]\rangle+\partial_{2}\langle X,D_{1}X\rangle-\partial_{1}\langle X,D_{2}X\rangle\nonumber \\
 &  & =-ig\langle B,[X,X^{\dagger}]\rangle+\partial_{2}\langle X,D_{1}X\rangle-\partial_{1}\langle X,D_{2}X\rangle\;.
\end{eqnarray}
Therefore, 
\begin{eqnarray}
\lefteqn{\langle D_{i}X,D_{i}X\rangle=}\nonumber \\
 &  & =\langle DX,DX\rangle+g\langle B,[X,X^{\dagger}]\rangle+\partial_{2}\langle X,iD_{1}X\rangle-\partial_{1}\langle X,iD_{2}X\rangle\;,\label{pI}
\end{eqnarray}
and similarly, 
\begin{eqnarray}
\lefteqn{\langle D_{i}X,D_{i}X\rangle=}\nonumber \\
 &  & =\langle\bar{D}X,\bar{D}X\rangle-g\langle B,[X,X^{\dagger}]\rangle-\partial_{2}\langle X,iD_{1}X\rangle+\partial_{1}\langle X,iD_{2}X\rangle\;,\label{pIbar}
\end{eqnarray}
\begin{equation}
\bar{D}=D_{1}-iD_{2}\;.
\end{equation}

\subsection{$SU(2)$ model}

\label{Bsu2}

For $SU(2)$, there is simply a one component positive root, $\alpha_{1}=\frac{1}{\sqrt{2}}$,
and $A_{11}=\frac{1}{2}$, $A^{11}=2$. Naming $\phi_{1}=\phi$, $\zeta_{1}=\zeta$,
the model in eq. (\ref{energy}) is, 
\begin{equation}
E=\int d^{2}x\,\left(\frac{1}{2}\langle B\rangle^{2}+\langle D_{i}\phi\rangle^{2}+\langle D_{i}\zeta\rangle^{2}+V_{{\rm Higgs}}-\langle\eta,B\rangle\right)\;.\label{ESU2}
\end{equation}
The Higgs potential can be written as, 
\begin{eqnarray}
V_{{\rm Higgs}}=\frac{1}{2}\langle\Phi\rangle^{2}+\langle Z\rangle^{2}\;,
\end{eqnarray}
\begin{equation}
\Phi=\sqrt{\gamma}\left(v_{c}\,\phi-[\zeta,\zeta^{\dagger}]\right)\makebox[.5in]{,}Z=\sqrt{\gamma_{z}}\left(v_{c}\,\zeta-2\,[\phi,\zeta]\right)\;.\label{ZSU2}
\end{equation}
Now, using 
\begin{eqnarray}
 &  & \langle B\rangle^{2}+\langle\Phi\rangle^{2}=\langle\Phi-B\rangle^{2}+2\langle\Phi,B\rangle\;.
\end{eqnarray}
and the property (\ref{pI}), for $X=\zeta$, namely, 
\begin{equation}
\langle D_{i}\zeta\rangle^{2}=\langle D\zeta\rangle^{2}+g\langle[\zeta,\zeta^{\dagger}],B\rangle+\partial_{2}\langle\zeta,iD_{1}\zeta\rangle-\partial_{1}\langle\zeta,iD_{2}\zeta\rangle\;,
\end{equation}
we obtain, 
\begin{eqnarray}
E & = & \int d^{2}x\,\left(\langle D_{i}\phi\rangle^{2}+\langle D\zeta\rangle^{2}+\langle Z\rangle^{2}\right.\nonumber \\
 &  & \left.+\frac{1}{2}\langle\Phi-B\rangle^{2}+\langle\Phi+g\,[\zeta,\zeta^{\dagger}]-\eta\,,B\rangle\right)\;.
\end{eqnarray}
Here, we have used a boundary condition at $(x^{1},x^{2})\to\infty$,
\begin{equation}
D_{i}\zeta\to0\;.\makebox[.5in]{{\rm for}}(x^{1},x^{2})\to\infty\;.
\end{equation}

Then, at $\gamma=g^{2}$ and taking the Lie algebra element, 
\begin{equation}
\eta=gv_{c}\,\phi\;,
\end{equation}
we get, 
\begin{eqnarray}
E & = & \int d^{2}x\,\left(\langle D_{i}\phi\rangle^{2}+\langle D\zeta\rangle^{2}+\langle Z\rangle^{2}+\frac{1}{2}\langle\Phi-B\rangle^{2}\right)\;.\nonumber \\
\end{eqnarray}
The bound is saturated when, 
\begin{equation}
D_{i}\phi=0\;,\label{q23}
\end{equation}
\begin{equation}
v_{c}\,\zeta-2\,[\phi,\zeta]=0\;,\label{withoutF}
\end{equation}

\begin{equation}
D\zeta=0\;,\label{mas}
\end{equation}
\begin{equation}
B=g\left(v_{c}\,\phi-[\zeta,\zeta^{\dagger}]\right)\;.\label{pe1}
\end{equation}
At the critical point, and taking $\gamma_{z}=g^{2}/2$, we can write,
\begin{eqnarray}
V_{{\rm Higgs}}=\frac{g^{2}}{2}\left[\langle v_{c}\,\phi-[\zeta,\zeta^{\dagger}]\rangle^{2}+\langle v_{c}\,\zeta-2\,[\phi,\zeta]\rangle^{2}\right]\;.
\end{eqnarray}
Using $\phi=\alpha_{1}\psi_{1}=\frac{1}{\sqrt{2}}\,\psi_{1}$, and
defining, 
\begin{equation}
\psi_{2}=\frac{\zeta+\zeta^{\dagger}}{\sqrt{2}}\makebox[.5in]{,}\psi_{3}=\frac{\zeta-\zeta^{\dagger}}{\sqrt{2}i}\makebox[.5in]{,}\sigma=\frac{v_{c}}{\sqrt{2}}\;,\label{p12}
\end{equation}
the model accepting BPS solutions is given by, 
\begin{equation}
E=\int d^{2}x\,\left(\frac{1}{2}\langle B\rangle^{2}+\frac{1}{2}\langle D_{i}\psi_{A}\rangle^{2}+V_{{\rm Higgs}}-gv_{c}\,\langle\phi,B\rangle\right)\;.\label{m1}
\end{equation}

\begin{eqnarray}
V_{{\rm Higgs}}=\frac{g^{2}}{2}\left[\langle\sigma\,\psi_{1}+i[\psi_{2},\psi_{3}]\rangle^{2}+\langle\sigma\,\psi_{2}+i[\psi_{3},\psi_{1}]\rangle^{2}+\langle\sigma\,\psi_{3}+i[\psi_{1},\psi_{2}]\rangle^{2}\right]\;,\nonumber \\
\end{eqnarray}
which is a modified version of the flavour symmetric model in eqs.
(\ref{mflavour}), (\ref{Vf}).

\subsection{$SU(3)$ model}

\label{Bsu3}

The Higgs potential is, 
\begin{equation}
V_{{\rm Higgs}}=\frac{1}{2}\langle\Phi,\Phi\rangle+\langle Z_{q},Z_{q}\rangle+{\cal R}\;,\label{H3}
\end{equation}
\begin{equation}
\Phi=s_{1}\,\Phi_{1}+s_{2}\,\Phi_{2}\makebox[.5in]{,}\Phi_{q}=\sqrt{\gamma}\left(v_{c}\,\phi_{q}-[\zeta_{(q)},\zeta_{(q)}^{\dagger}]\right)\;,
\end{equation}
\begin{equation}
{\cal R}=\gamma_{r}\left(\langle[\phi_{1},\phi_{2}]\rangle^{2}+\langle[\phi^{1},\zeta_{2}]\rangle^{2}+\langle[\phi^{2},\zeta_{1}]\rangle^{2}\right)\;.
\end{equation}
To obtain a set of BPS equations, we initially diagonalize the $\phi_{q}$-kinetic
term in eq. (\ref{energy}). Note that any quantity of the form $\langle X_{q},X^{q}\rangle$
can be written as, 
\begin{equation}
\langle X_{q},X^{q}\rangle=\frac{1}{2}\langle X_{2}+X_{1},X^{2}+X^{1}\rangle+\frac{1}{2}\langle X_{2}-X_{1},X^{2}-X^{1}\rangle\;.
\end{equation}
On the other hand, the Cartan matrix for $SU(3)$ is, 
\begin{equation}
\mathbb{C}=6\,\mathbb{A}=\left(\begin{array}{rr}
2 & -1\\
-1 & 2
\end{array}\right)\;,
\end{equation}
$\mathbb{C}|_{qp}=C_{qp}$, $\mathbb{A}|_{qp}=A_{qp}$. Therefore,
\begin{equation}
X_{2}+X_{1}=(A_{22}+A_{12})\, X^{2}+(A_{21}+A_{11})\, X^{1}=\frac{1}{6}\,(X^{2}+X^{1})\;,
\end{equation}
\begin{equation}
X_{2}-X_{1}=(A_{22}-A_{12})\, X^{2}+(A_{21}-A_{11})\, X^{1}=\frac{1}{2}\,(X^{2}-X^{1})\;.
\end{equation}
That is, 
\begin{equation}
\langle X_{q},X^{q}\rangle=3\,\langle X_{+},X_{+}\rangle+\langle X_{-},X_{-}\rangle\makebox[.3in]{,}X_{+}=X_{2}+X_{1}\makebox[.3in]{,}X_{-}=X_{2}-X_{1}\;,
\end{equation}
and the energy functional in eq. (\ref{energy}) results, 
\begin{eqnarray}
E & = & \int d^{2}x\,\left(\frac{1}{2}\langle B\rangle^{2}+\frac{3}{2}\langle D_{i}\phi_{+}\rangle^{2}+\frac{1}{2}\langle D_{i}\phi_{-}\rangle^{2}+\langle D_{i}\zeta_{q},D_{i}\zeta_{q}\rangle\right)\nonumber \\
 &  & +\int d^{2}x\,\left(\frac{1}{2}\langle\Phi\rangle^{2}+\langle Z_{q},Z_{q}\rangle+{\cal R}-\langle\eta,B\rangle\right)\;.\label{ene3}
\end{eqnarray}
Next, similarly to the $SU(2)$ case, using, 
\begin{eqnarray}
 &  & \langle B\rangle^{2}+\langle\Phi\rangle^{2}=\langle\Phi-B\rangle^{2}+2\langle\Phi,B\rangle\;,
\end{eqnarray}
and properties (\ref{pI}), (\ref{pIbar}) for $X=\zeta_{2}$,
$\zeta_{1}$, respectively, 
\[
\langle D_{i}\zeta_{2},D_{i}\zeta_{2}\rangle=\langle D\zeta_{2},D\zeta_{2}\rangle+g\langle[\zeta_{2},\zeta_{2}^{\dagger}],B\rangle+\partial_{3}\langle\zeta_{2},iD_{2}\zeta_{2}\rangle-\partial_{2}\langle\zeta_{2},iD_{3}\zeta_{2}\rangle\;,
\]
\[
\langle D_{i}\zeta_{1},D_{i}\zeta_{1}\rangle=\langle\bar{D}\zeta_{1},\bar{D}\zeta_{1}\rangle-g\langle[\zeta_{1},\zeta_{1}^{\dagger}],B\rangle-\partial_{3}\langle\zeta_{1},iD_{2}\zeta_{1}\rangle+\partial_{2}\langle\zeta_{1},iD_{3}\zeta_{1}\rangle\;,
\]
we obtain, 
\begin{eqnarray}
E & = & \int d^{2}x\,\left(\frac{3}{2}\langle D_{i}\phi_{+}\rangle^{2}+\frac{1}{2}\langle D_{i}\phi_{-}\rangle^{2}+\langle\bar{D}\zeta_{1}\rangle^{2}+\langle D\zeta_{2}\rangle^{2}+\langle Z_{q},Z_{q}\rangle\right)\nonumber \\
 &  & +\int d^{2}x\,\left(\frac{1}{2}\langle\Phi-B\rangle^{2}+\langle\Phi+g[\zeta_{2},\zeta_{2}^{\dagger}]-g[\zeta_{1},\zeta_{1}^{\dagger}]-\eta\,,B\rangle\right)\nonumber \\
 &  & +\int d^{2}x\,\gamma_{r}\left(\langle[\phi_{1},\phi_{2}]\rangle^{2}+\langle[\phi^{1},\zeta_{2}]\rangle^{2}+\langle[\phi^{2},\zeta_{1}]\rangle^{2}\right)\;,
\end{eqnarray}
where we have used that the system is in a local vacuum at $(x^{1},x^{2})\to\infty$.

Then, taking $s_{1}=-1$, $s_{2}=1$, $\gamma=g^{2}$, which gives,
\begin{equation}
\Phi=gv_{c}\,\phi_{-}-g\,\left([\zeta_{2},\zeta_{2}^{\dagger}]-[\zeta_{1},\zeta_{1}^{\dagger}]\right)\makebox[.5in]{,}\phi_{-}=\phi_{2}-\phi_{1}\;,\label{Phi-}
\end{equation}
and the Lie algebra element, 
\begin{equation}
\eta=gv_{c}\,\phi_{-}\;,
\end{equation}
we obtain, 
\begin{eqnarray*}
E & = & \int d^{2}x\,\left(\frac{3}{2}\langle D_{i}\phi_{+}\rangle^{2}+\frac{1}{2}\langle D_{i}\phi_{-}\rangle^{2}+\langle\bar{D}\zeta_{1}\rangle^{2}+\langle D\zeta_{2}\rangle^{2}+\langle Z_{q},Z_{q}\rangle\right)\\
 &  & +\int d^{2}x\,\left(\frac{1}{2}\langle\Phi-B\rangle^{2}+\gamma_{r}\left(\langle[\phi_{1},\phi_{2}]\rangle^{2}+\langle[\phi^{1},\zeta_{2}]\rangle^{2}+\langle[\phi^{2},\zeta_{1}]\rangle^{2}\right)\right)\;.
\end{eqnarray*}
Therefore, the BPS equations are, 
\begin{equation}
D_{i}\phi_{-}=0\makebox[.5in]{,}D_{i}\phi_{+}=0\;,\label{q23n}
\end{equation}
\begin{equation}
v_{c}\,\zeta_{2}-[\phi^{2},\zeta_{2}]=0\makebox[.5in]{,}\phi^{2}=3\phi_{+}+\phi_{-}\;,\label{withoutF1}
\end{equation}
\begin{equation}
v_{c}\,\zeta_{1}-[\phi^{1},\zeta_{1}]=0\makebox[.5in]{,}\phi^{1}=3\phi_{+}-\phi_{-}\;,
\end{equation}
\begin{equation}
[\phi_{1},\phi_{2}]=0\makebox[.3in]{,}[\phi^{1},\zeta_{2}]=0\makebox[.3in]{,}[\phi^{2},\zeta_{1}]=0\;.\label{tr}
\end{equation}

\begin{equation}
D\zeta_{2}=0\makebox[.5in]{,}\bar{D}\zeta_{1}=0\;,\label{mas1}
\end{equation}
\begin{equation}
B=g\left(v_{c}\,\phi_{-}-\left([\zeta_{2},\zeta_{2}^{\dagger}]-[\zeta_{1},\zeta_{1}^{\dagger}]\right)\right)\;.\label{pe1n}
\end{equation}

\section{Center vortex ansatz}

In order to propose a center vortex ansatz, it would be useful having a parametrization analogous to the simple $U(1)$ case, where evidencing the modulus and the phase of the complex Higgs field, $\rho\, e^{i\chi}$, accompanied by the
gauge field, $a\,\partial_{i}\chi$, permits the implementation of
boundary conditions. For this purpose, we could initially determine whether the asymptotic vacua
are of the form given in eq. (\ref{defi}) or (\ref{conju}),
and then look for the mapping $S$ (the non Abelian phase) such that,
\begin{equation}
S(\vec{\alpha}_{q}\cdot\vec{H})S^{-1}\makebox[.5in]{,}SE_{\alpha_{q}}S^{-1}\;,\label{locb}
\end{equation}
respectively
\begin{equation}
S(-\vec{\alpha}_{q}\cdot\vec{H})S^{-1}\makebox[.5in]{,}S(-E_{\alpha_{q}}^{T})S^{-1}\;,\label{locb1}
\end{equation}
is the closest local basis to the field configuration $\phi_{q}$,
$\zeta_{q}$. The ``polar'' decomposition is then, 
\begin{equation}
\phi_{q}=SF_{q}S^{-1}\makebox[.5in]{,}\zeta_{q}=SZ_{q}S^{-1}\;.\label{cansatz}
\end{equation}

The notion of closest mapping can be obtained by following similar
steps to those used when defining adjoint Laplacian center gauges
\cite{deFP}. For example, in $SU(2)$, we can take $\psi_{1}=\sqrt{2}\,\phi_{1}$,
together with $\psi_{2}$, $\psi_{3}$ (obtained from $\zeta_{1}$
using eq. (\ref{p12})), and expand these fields in the $T_{A}$ basis,
\begin{equation}
\psi_{A}=\psi_{AB}\, T_{B}\makebox[.5in]{,}\psi_{AB}=\langle\psi_{A},T_{B}\rangle\;,
\end{equation}
$A=1,2,3$. The real elements $\psi_{AB}$ form a $3\times3$ matrix
$\Psi$, for which a polar decomposition exists, 
\begin{equation}
\Psi=Q\, R\;,
\end{equation}
where $R\in SO(3)$ and $Q$ is real symmetric and positive semidefinite.
The closest orthogonal matrix to $\Psi$ is $R$, then the closest
orthonormal basis to $\{\psi_{A}\}$ is given by, 
\begin{equation}
n_{A}=R_{AB}\, T_{B}=ST_{A}S^{-1}\;,\label{lf}
\end{equation}
where $S$ is defined up to a global center element. That is, for
$SU(2)$ adjoint Higgs fields, the ``modulus and phase'' decomposition
is, 
\begin{equation}
\psi_{A}=S(Q_{AB}\, T_{B})S^{-1}\;,
\end{equation}
which can be translated back to $\phi_{1}$, $\zeta_{1}$-language.

With regard to the gauge field, we note that on any simply connected
region, which does not contain the pointlike defects of the local
basis, the Higgs field ansatz looks as a gauge transformation. Therefore,
in that region, the field equations would be simplified by representing
the smooth $A_{i}$ as a gauge transformation of a vector field ${\cal A}_{i}$.
However, in the defining representation, $S$ is in general discontinuous
on some curves, as it changes by a center element when we go around
a center vortex. Therefore, on $R^{2}-\{{\rm pointlike~defects}\}$,
the ansatz, 
\begin{equation}
S{\cal A}_{i}S^{-1}+\frac{i}{g}\, S\partial_{i}S^{-1}\;,
\end{equation}
cannot work, as it contains a contribution ($I_{i}$) concentrated
at the points where $S^{-1}$ is discontinuous. There are three equivalent
possibilities to circumvent this problem. 
\begin{itemize}
\item proceed as in ref. \cite{engelhardt1,reinhardt}, proposing the parametrization,
\begin{equation}
A_{i}=S{\cal A}_{i}S^{-1}+\frac{i}{g}\, S\partial_{i}S^{-1}-I_{i}\;,\label{an1}
\end{equation}

\item proceed as in \cite{lucho,Lucho2}, to write 
\begin{equation}
A_{i}=({\cal A}_{i}^{A}-C_{i}^{A})\, n_{A}\;,\label{an2}
\end{equation}
where $C_{i}^{A}$ only depends on the local colour frame (\ref{lf}). 
\item work with the fields mapped into the adjoint representation\footnote{The matrices $M_{A}$ are generators of the adjoint representation. },
\begin{equation} 
Ad(A_{i})=RAd({\cal A}_{i})R^{-1}+\frac{i}{g}R\partial_{i}R^{-1}\makebox[.5in]{,}Ad({\cal A}_{i})={\cal A}_{i}^{A}\, M_{A}\;,\label{gad}
\end{equation}
\begin{equation}
Ad(\psi_{q})=RAd(P_{q})R^{-1}\makebox[.5in]{,}Ad(\zeta_{\alpha})=RAd(P_{\alpha})R^{-1}\;,\label{anad}
\end{equation}

\end{itemize}
where we used, 
\begin{equation}
R_{AB}\, M_{B}=RM_{A}R^{-1}\;.
\end{equation}
Here, we shall use the third possibility. The advantage of the second
and third options is that $n_{A}$ and $R$ contain at most pointlike
defects, as they are always single-valued when we go around a loop.
Then, the $R\partial_{i}R^{-1}$ term does not introduce delta distributions
concentrated on curves, and a smooth $Ad(A_{i})$ ansatz can be implemented
with $Ad({\cal A}_{i})$ satisfying appropriate boundary conditions
at the vortex guiding centers.

It is important to underline that, in the ansatz (\ref{gad}), $Ad(A_{i})$
\textit{is not } a gauge transformation of $Ad({\cal A}_{i})$. The
magnetic field $B$ is given by, 
\begin{equation}
Ad(B)=RAd({\cal B})R^{-1}+\frac{i}{g}R[\partial_{1},\partial_{2}]R^{-1}\;,\label{adja1}
\end{equation}
\begin{equation}
{\cal B}=\partial_{1}{\cal A}_{2}-\partial_{2}{\cal A}_{1}-ig[{\cal A}_{1},{\cal A}_{2}]\;,\label{field-st}
\end{equation}
where the last term in eq. (\ref{adja1}) is concentrated at the vortex
guiding centers. The profiles ${\cal A}_{i}$, $F_{q}$ and $Z_{q}$,
must be such that $B$ and the Higgs fields be well-defined and smooth
everywhere, and satisfy the desired asymptotic behaviour.
For a single center vortex, with charge $z$ modulo $N$,  we can impose in the asymptotic
region, 
\begin{equation}
{\cal A}_{i}\to0\makebox[.5in]{,}S\to e^{i\varphi\,\vec{\beta}\cdot\vec{T}}\;,
\end{equation}
where $\vec{\beta}$ satisfies eq. (\ref{poss-b}). When minimizing
the energy, the extension of $R=Ad(S)$ from the asymptotic region
to the vortex core should not only contemplate keeping $R$ along
a Cartan direction but also other possibilities. In this regard, note
that for 
\begin{equation}
\vec{\beta}-\vec{\beta}_{0}=2N\vec{\gamma}\makebox[.5in]{,}\vec{\gamma}\in\Lambda(Ad(SU(N)))\;,\label{c0}
\end{equation}
where $\Lambda(\dots)$ represents the lattice of weights of the adjoint
representation (or root lattice), it is always possible to obtain a map $R(r,\varphi)$ verifying, 
\begin{equation}
R(r,\varphi)=\left\{ \begin{array}{ll}
e^{i\varphi\,\vec{\beta}\cdot\vec{M}}, & r>r_{m}\\
e^{i\varphi\,\vec{\beta}_{0}\cdot\vec{M}}, & r<r_{0}\;,
\end{array}\right.\label{beha}
\end{equation}
that is smooth for $r\geq r_{0}$. This map can be constructed as
$R=e^{i\varphi\,\vec{\beta}_{0}\cdot\vec{M}}R_{0}$, with 
\begin{equation}
R_{0}(r,\varphi)=\left\{ \begin{array}{ll}
e^{i\varphi\,2N\vec{\gamma}\cdot\vec{M}}, & r>r_{m}\\
I, & r<r_{0}\;.
\end{array}\right.
\end{equation}
Note that $R_0$ always exists as $e^{i\varphi\,2N\vec{\gamma}\cdot\vec{T}}$ is a
closed path in $SU(N)$, and therefore $e^{i\varphi\,2N\vec{\gamma}\cdot\vec{M}}$
is topologically trivial in $Ad(SU(N))$. Different magnetic weights $\vec{\beta}_{0}$
imply different types of defect and profile function behaviours at
$r\to0$. For example, an asymptotic behaviour with $z=0$  is described
by any $\vec{\beta}\in2N\,\Lambda(Ad(SU(N)))$. All these values can
be extended to $\vec{\beta}_{0}=0$. For this choice, $R(r,\varphi)$
contains no defect at the origin and the minimization process will
simply return a trivial result, corresponding to a pure gauge transformation
of the vacuum configuration. For $z=\pm1$, i.e. $\vec{\beta}=2N\,\vec{w}+2N\,\vec{\gamma}$, where $\vec{w}$ is a weight of the defining representation or its conjugate, there is no manner to avoid a defect at $r\to0$. The energy is expected to be minimized by $\vec{\beta}_{0}=2N\,\vec{w}$, as in this case
some of the basis components will only give one turn when we go around
a small circle centered at $r=0$.

\section{BPS center vortices}

At the critical point, to solve the $SU(2)$ and $SU(3)$ BPS equations,
it will be enough to consider, 
\begin{equation}
\phi_{q}=v_{c}\, S(\vec{\alpha}_{q}\cdot\vec{H})S^{-1}\makebox[.5in]{,}\zeta_{q}=u\, SE_{\alpha_{q}}S^{-1}\;,
\end{equation}
(and a similar expression for the conjugate sector). The possible
non Abelian phases $S$ are such that $R=Ad(S)$ behaves as in eq.
(\ref{beha}). As we will see, the $A_{i}$ parametrization in terms
of ${\cal A}_{i}$ and $S$ together with the BPS equations imply,
\begin{equation}
{\cal A}_{i}=c_{i}\,\vec{\delta}\cdot\vec{H}\;, 
\end{equation}
where $H_{q}$ denote the Cartan generators. Then, from eqs.
(\ref{an1})-(\ref{gad}), for $r<r_{0}$ the gauge field is, 
\begin{equation}
A_{i}=c_{i}\,\vec{\delta}\cdot\vec{H}+\frac{1}{g}\partial_{i}\varphi\,\vec{\beta}_{0}\cdot\vec{H}\;,\label{beh0}
\end{equation}
and in order to obtain a regular magnetic field, we must have $\vec{\delta}=\pm\vec{\beta}_{0}$
and $c_{i}\to\mp\frac{1}{g}\partial_{i}\varphi$, when $r\to0$.

\subsection{$\mathfrak{su}(2)$}

For nonzero $\zeta$, eq. (\ref{withoutF}) implies, 
\begin{equation}
\phi=\frac{v_{c}}{\sqrt{2}}\, SH_{1}S^{-1}\makebox[.5in]{,}\zeta=u\, SE_{\alpha_{1}}S^{-1}\;,
\end{equation}
where the possible magnetic weights $\beta$ in eq. (\ref{beha}) are $\beta=q\sqrt{2}$,
$q\in\mathbb{Z}$. Now, using any of the parametrizations (\ref{an1})-(\ref{gad}),
eq. (\ref{q23}) gives, 
\begin{equation}
D_{i}({\cal A})(H_{1})=0\makebox[.9in]{{\rm or},}[{\cal A}_{i},H_{1}]=0\;,
\end{equation}
whose solution is, 
\begin{equation}
{\cal A}_{i}=c_{i}\,\beta_{0}H_{1}\;,
\end{equation}
(the case $\delta=-\beta_{0}$ is discussed at the end). Similarly,
eq. (\ref{mas}) becomes, 
\begin{equation}
D({\cal A})(uE_{\alpha_{1}})=0\makebox[.9in]{{\rm or},}(\partial_{1}+i\partial_{2})u\, E_{\alpha_{1}}-igu\,[{\cal A}_{1}+i{\cal A}_{2},E_{\alpha_{1}}]=0\;.
\end{equation}
Thus, joining this information, we obtain, 
\begin{equation}
\frac{\beta_{0}}{\sqrt{2}}\,(c_{1}+ic_{2})=\frac{1}{2g}\,(\partial_{2}h-i\partial_{1}h)\makebox[.5in]{,}u=v_{c}\, e^{h/2}\;.
\end{equation}
\begin{equation}
Ad({\cal B})=(\partial_{1}c_{2}-\partial_{2}c_{1})\,\beta_{0}Ad(H_{1})=-\frac{1}{\sqrt{2}g}(\partial_{1}^{2}+\partial_{2}^{2})h\, Ad(H_{1})\;,
\end{equation}
where we have changed the variables from $u$ to $h=2\ln(u/v_{c})$,
as is usually done in the $U(1)$ case. Therefore, eqs. (\ref{pe1}) and
(\ref{adja1}) imply, 
\begin{equation}
\left((\partial_{1}^{2}+\partial_{2}^{2})h+g^{2}v_{c}^{2}\,(1-e^{h})\right)RAd(H_{1})R^{-1}=i\sqrt{2}\, R[\partial_{1},\partial_{2}]R^{-1}\;.
\end{equation}
The second member is obtained from eq. (\ref{beha}), 
\begin{eqnarray}
i\, R[\partial_{1},\partial_{2}]R^{-1}=\beta_{0}\,[\partial_{1},\partial_{2}]\varphi\, Ad(H_{1})\;.
\end{eqnarray}
As is well-known, although for $j_{i}=\partial_{i}\varphi$ the quantity
$\partial_{2}j_{3}-\partial_{3}j_{2}=[\partial_{2},\partial_{3}]\varphi$
seems to vanish, it is in fact concentrated at $x^{2}=x^{3}=0$, where
$e^{i\varphi}$ contains a defect. Namely, 
\begin{equation}
\partial_{1}j_{2}-\partial_{2}j_{1}=2\pi\,\delta^{(2)}(x^{1},x^{2})\;.
\end{equation}
This can be checked using Stokes' theorem. Then, we get, 
\begin{equation}
(\partial_{1}^{2}+\partial_{2}^{2})\, h+g^{2}v_{c}^{2}\,(1-e^{h})=2\pi\sqrt{2}\beta_{0}\,\delta^{(2)}(x^{1},x^{2})\;.\label{anzeq}
\end{equation}
For $q$ even, the asymptotic behaviour $S=e^{iq\sqrt{2}\,\varphi H_{1}}$,
$R=e^{iq\sqrt{2}\,\varphi Ad(H_{1})}$, on the circle $r\to\infty$,
can be continuously changed to a behaviour characterized by $\beta_{0}=0$,
as $r$ is varied from $\infty$ to $0$. The abscence of defects
will lead to a trivial pure gauge solution for the BPS equations.
On the other hand, for $q$ odd, the asymptotic behavior can be changed
to $\beta_{0}=+\sqrt{2}$, as well as $\beta_{0}=-\sqrt{2}$. For
these values, the frame components $ST_{\alpha_1} S^{-1}$, $ST_{\bar{\alpha}_1}S^{-1}$,
\begin{equation}
T_{\alpha}=\frac{1}{\sqrt{2}}(E_{\alpha}+E_{-\alpha})\makebox[.5in]{,}T_{\bar{\alpha}}=\frac{1}{\sqrt{2}i}(E_{\alpha}-E_{-\alpha})\;,\label{Tes}
\end{equation}
rotate only once, when we go close to and around the origin. The solution
to eq. \eqref{anzeq} is well-defined for $\beta_{0}=+\sqrt{2}$,
while it is ill-defined for $\beta_{0}=-\sqrt{2}$. In the latter
case, the well-defined solution is obtained using the conjugate ansatz
(see a similar discussion in \ref{su3e}), 
\begin{equation}
\phi=\frac{v_{c}}{\sqrt{2}}\, S(-H_{1})S^{-1}\makebox[.5in]{,}\zeta=u\, S(-E_{\alpha_{1}}^{T})S^{-1}\makebox[.5in]{,}{\cal A}_{i}=c_{i}\,(-\sqrt{2})H_{1}\;.
\end{equation}
In $SU(2)$, both the vortex and its conjugate satisfy, 
\begin{equation}
S(\varphi+2\pi)=-S(\varphi)\;,
\end{equation}
so they are equivalent objects.

\subsection{$\mathfrak{su}(3)$}

\label{su3e}

The equations (\ref{withoutF1})-(\ref{tr}) imply,
\begin{equation}
\phi_{q}=v_{c}\, S(\vec{\alpha}_{q}\cdot\vec{H})S^{-1}\makebox[.5in]{,}\zeta_{q}=u_{q}\, SE_{\alpha_{q}}S^{-1}\label{ang}
\end{equation}
(q=1,2) so that imposing eq. (\ref{q23n}), we obtain, 
\begin{equation}
D_{i}({\cal A})(\vec{\alpha}_{q}\cdot\vec{H})=0\makebox[.9in]{{\rm or},}[{\cal A}_{i},\vec{\alpha}_{q}\cdot\vec{H}]=0\;.
\end{equation}
This means that ${\cal A}_{i}$ is in the Cartan subalgebra. Taking
$\vec{\delta}=+\vec{\beta}_{0}$, 
\begin{equation}
{\cal A}_{i}=c_{i}\,\vec{\beta}_{0}\cdot\vec{H}\;,
\end{equation}
eq. (\ref{mas1}) gives, 
\begin{eqnarray}
 &  & (\partial_{1}+i\partial_{2})u_{2}\, E_{\alpha_{2}}-ig\,(c_{1}+ic_{2})u_{2}\,[\vec{\beta}_{0}\cdot\vec{H},E_{\alpha_{2}}]=0\;,
\end{eqnarray}
\begin{eqnarray}
 &  & (\partial_{1}-i\partial_{2})u_{1}\, E_{\alpha_{1}}-ig\,(c_{1}-ic_{2})u_{1}\,[\vec{\beta}_{0}\cdot\vec{H},E_{\alpha_{1}}]=0\;.
\end{eqnarray}
Then, we get, 
\begin{equation}
(\vec{\beta}_{0}\cdot\vec{\alpha}_{2})\,(c_{1}+ic_{2})=\frac{1}{2g}\,(\partial_{2}h_{2}-i\partial_{1}h_{2})\;,\label{cm}
\end{equation}
\begin{equation}
(\vec{\beta}_{0}\cdot\vec{\alpha}_{1})\,(c_{1}-ic_{2})=-\frac{1}{2g}\,(\partial_{2}h_{1}+i\partial_{1}h_{1})\;,\label{cc}
\end{equation}
\begin{equation}
{\cal B}=-\frac{1}{2g}(\partial_{1}^{2}+\partial_{2}^{2})h\,\frac{\vec{\beta}_{0}\cdot\vec{H}}{\vec{\beta}_{0}\cdot\vec{\alpha}_{2}}\;,\label{3i}
\end{equation}
where $u_{i}=v_{c}\, e^{h_{i}/2}$. In addition, eq. (\ref{pe1n}) reads, 
\begin{eqnarray*}
\lefteqn{iR[\partial_{1},\partial_{2}]R^{-1}=}\\
 &  & =R\, Ad\left(-g\,{\cal B}+g^{2}v_{c}^{2}\,(\vec{\alpha}_{2}-\vec{\alpha}_{1})\cdot\vec{H}-g^{2}v_{c}^{2}\,(\vec{\alpha}_{2}\, e^{h_{2}}-\vec{\alpha}_{1}\, e^{h_{1}})\cdot\vec{H}\right)R^{-1}\;,
\end{eqnarray*}
while for a single vortex, eq. (\ref{beha}) implies, 
\begin{eqnarray}
i\, R[\partial_{1},\partial_{2}]R^{-1}=[\partial_{1},\partial_{2}]\varphi\, Ad(\vec{\beta}_{0}\cdot\vec{H})=2\pi\,\delta^{(2)}(x^{1},x^{2})\, Ad(\vec{\beta}_{0}\cdot\vec{H})\;.
\end{eqnarray}
Putting this information together, we arrive at, 
\begin{equation}
-g\,{\cal B}+g^{2}v_{c}^{2}\,(\vec{\alpha}_{2}-\vec{\alpha_{1}})-g^{2}v_{c}^{2}\,(\vec{\alpha}_{2}\, e^{h_{2}}-\vec{\alpha}_{1}\, e^{h_{1}})=2\pi\,\delta^{(2)}(x^{1},x^{2})\,\vec{\beta}_{0}\;.\label{pt}
\end{equation}
Let us consider the case where $\vec{\beta}_{0}$ is associated with
a weight of the defining representation. Noting that ${\cal B}=(\partial_{1}c_{2}-\partial_{2}c_{1})\,\vec{\beta}_{0}\cdot\vec{H}$
and $\vec{\alpha}_{2}-\vec{\alpha}_{1}=\frac{1}{2}\,\vec{\beta}_{2}$,
in order to have a nontrivial solution we are led to $\vec{\beta}_{0}=\pm\vec{\beta}_{2}$.
For these cases, $\vec{\beta}_{0}\cdot\vec{\alpha}_{2}=-\vec{\beta}_{0}\cdot\vec{\alpha}_{1}$, the equations (\ref{cm}) and (\ref{cc}) give $h_{1}=h_{2}=h$, $u_{i}=u=e^{h}$, and both sides of eq. (\ref{pt}) turn out to be oriented along the
same direction. Under these conditions, we obtain, 
\begin{equation}
(\vec{\beta}_{0}\cdot\vec{\alpha}_{2})^{-1}\,(\partial_{1}^{2}+\partial_{2}^{2})h\,\vec{\beta}_{0}+g^{2}v_{c}^{2}\,(1-e^{h})\,\vec{\beta}_{2}=4\pi\,\delta^{(2)}(x^{1},x^{2})\,\vec{\beta}_{0}\;.\label{3ii}
\end{equation}
That is, for $\vec{\beta}_{0}=+\vec{\beta}_{2}$ ($\vec{\beta}_{0}\cdot\vec{\alpha}_{2}=+1$),
\begin{equation}
(\partial_{1}^{2}+\partial_{2}^{2})h+g^{2}v_{c}^{2}\,(1-e^{h})=4\pi\,\delta^{(2)}(x^{1},x^{2})\;.\label{well}
\end{equation}
On the other hand, the choice $\vec{\beta}_{0}=-\vec{\beta}_{2}$
would imply, 
\begin{equation}
(\partial_{1}^{2}+\partial_{2}^{2})h+g^{2}v_{c}^{2}\,(1-e^{h})=-4\pi\,\delta^{(2)}(x^{1},x^{2})\;.\label{ill}
\end{equation}
The second choice does not lead to well-defined Higgs fields. In effect,
while close to the origin eq. (\ref{well}) gives $h\sim2\ln r$,
$u=e^{h/2}\sim r$, producing single-valued Higgs fields (and $c_{i}\sim-\frac{1}{g}\,\partial_{i}\varphi$),
eq. (\ref{ill}) gives $h\sim-2\ln r$, $u=e^{h/2}\sim1/r$. However,
it is easy to see that the new ansatz obtained from (\ref{ang}) by
the replacement, 
\begin{equation}
\vec{\alpha}_{q}\cdot\vec{H}\to-\vec{\alpha}_{q}\cdot\vec{H}\makebox[.5in]{,}E_{\alpha_{q}}\to-E_{\alpha_{q}}^{T}\;,\label{conjug}
\end{equation}
solves the BPS equations with a well-defined $h$ satisfying eq.
(\ref{well}), provided we choose $\vec{\beta}_{0}=-\vec{\beta}_{2}$.
Other weights can be obtained by replacing in eq. (\ref{ang}) (resp.
eq. (\ref{conjug})), 
\begin{equation}
\vec{\alpha}_{q}\to\vec{\alpha}_{q}^{\, W}\;,
\end{equation}
where $W$ is a Weyl transformation. The solutions will be characterized
by the gauge field behaviour (\ref{beha}), with $\vec{\beta}_{0}\to\vec{\beta}_{0}^{\, W}=\vec{\beta}_{2}^{\, W}$
(resp. $\vec{\beta}_{0}\to\vec{\beta}_{0}^{\, W}=-\vec{\beta}_{2}^{\, W}$)
(and $\vec{\beta}\to\vec{\beta}^{\, W}$). Then, these solutions are
characterized by the weights of the defining representation, $\vec{\beta}_{1},\vec{\beta}_{2},\vec{\beta}_{3}$,
and their conjugates, $-\vec{\beta}_{1},-\vec{\beta}_{2},-\vec{\beta}_{3}$.
As the mappings $S$ satisfy, 
\begin{equation}
S(\varphi+2\pi)=e^{\pm i2\pi/3}\, S(\varphi)\;,
\end{equation}
they correspond to center vortices with the minimum charges $z=\pm1$.

\section{Conclusions}

In this article we presented Yang-Mills-Higgs nonrelativistic models with $SU(N)\to Z(N)$
SSB pattern that accept BPS center vortex equations (for $N=2,3$).

For this purpose, we initially proposed a class of $SU(N)$ Lorentz invariant models
containing real and complex adjoint Higgs fields, that can be labelled
by the simple roots of the $\mathfrak{su}(N)$ Lie algebra. The Higgs
potential is such that its minimization returns a set of conditions
that essentially define a Chevalley basis. The space of vacua also
contains a trivial symmetry preserving point, where the Higgs fields
vanish, separated from the SSB points by a potential barrier.

Next, we introduced a nonrelativistic interaction term so as to obtain
a set of BPS equations. This is a term that tends to align, in the
Lie algebra, the magnetic field and one of the Higgs fields. Finally,
we obtained some solutions. For example, the $Z(3)$ vortices come
in three colours (the weights of the defining representation), which
are physically equivalent, and three anticolours, obtained from an
ansatz based on the conjugate representation.

Generally, BPS equations are derived by working on the energy functional,
which is a local object, and obtaining a bound that only depends on
some topological charge. For $U(1)$ vortices, the bound is given
by the magnetic flux. This is a topological term that can be written
locally, by means of a flux density. On the other hand, for center
vortices, the flux concept is given by the asymptotic behaviour of
the gauge invariant Wilson loop, a nonlocal object that may not arise
in the calculation. For this reason, the search for BPS equations
led us to consider the alignment interaction. After completing the
squares, the energy is always greater than or equal to zero. Thus,
BPS center vortices are characterized by an exact compensation between
the positive definite part of the energy functional (kinetic energy
plus Higgs potential) and the contribution originated from alignment.

Similarly to the minima of the Higgs potential, the BPS equations
have trivial solutions with vanishing Higgs fields (and pure gauge
fields) and a sector where the asymptotic fields are in SSB vacua.
Although the BPS solutions have vanishing energy, no finite energy 
configurations continuously interpolating the center vortex ($z=\pm1$) 
and the trivial configuration ($z=0$) exist. In other words, there is an energy barrier for the continuous deformation of one configuration into the other. The general solution to the BPS equations was written in terms of a reduced set
of profile functions and a mapping $R(S)$ in the adjoint representation
of $SU(N)$. The mapping $S\in SU(N)$, contains information about
the asymptotic Wilson loop and the set of possible defects at the
vortex guiding centers, which determine the behaviour of the profile
functions.

In spite of the Abelian looking profile functions obtained, we would
like to underline two important differences. As the number of BPS
center vortices is increased, the energy continues to vanish. This is
in contrast to the $U(1)$ case, where the energy increases linearly
with the number of vortices, a property that is modified below and
above the critical coupling, implying either attractive or repulsive
forces. In addition, the topological properties of the adjoint representation
of $SU(N)$ modify the relation between asymptotic phases and defects.
A $U(1)$ asymptotic phase implies a unique type of pointlike defect,
and a unique order for the zero of the corresponding Higgs profile
function. On the other hand, for an asymptotic non Abelian phase,
many extensions to reach a pointlike defect are possible, with corresponding
conditions on the profile functions. For $Z(N)$-charge equal to $\pm1$,
a defect is always present, while for vanishing $Z(N)$-charge the
defects in $R(S)$ can be avoided, and the lowest energy solution
simply corresponds to a regular gauge transformation of the SSB vacua.
When leaving the critical point, the new energetics, topology,
and field content are expected to modify the forces between center
vortices, as compared with the $U(1)$ case. This may be a possibility
worth exploring. 

Summarizing, the search for BPS bounds is among the preferred analytical tools to understand topological objects. In this manuscript, we showed what would be the situation in the context of center vortex $(2+1)$d models: they become nonrelativistic. Then, although BPS center vortices are not directly physically relevant, they could provide a useful concept when embarking on numerical simulations. The presence of a BPS point in the extended parameter space could serve as a check of the numerical analysis when moving away from the physically relevant non BPS Lorentz invariant confining models. 

\section*{Acknowledgements}

The Conselho Nacional de Desenvolvimento Científico e Tecnológico
(CNPq-Brazil), the Fundação de Amparo a Pesquisa do Estado do Rio
de Janeiro (FAPERJ), and the Proppi-UFF are acknowledged for their
financial support.

\section*{Appendix A: Cartan decomposition of $\mathfrak{g}$}

A compact connected simple Lie algebra $\mathfrak{g}$ can be decomposed
in terms of hermitian Cartan generators $H_{q}$, $q=1,\dots,r$,
which generate a Cartan subgroup $H$, and off-diagonal generators
$E_{\alpha}$, or root vectors. The latter are labelled by a system of roots $\vec{\alpha}=(\alpha_{1},\dots,\alpha_{r})$.
They satisfy, 
\begin{equation}
[H_{q},H_{p}]=0\makebox[.3in]{,}[H_{q},E_{\alpha}]=\alpha_{q}\, E_{\alpha}\makebox[.3in]{,}[E_{\alpha},E_{\alpha}^{\dagger}]=\langle E_{\alpha},E_{\alpha}\rangle\, H_{\alpha}\;,\label{algebrag}
\end{equation}
where, for every root $\vec{\alpha}$, $H_{\alpha}$ is defined by,
\begin{equation}
\langle H_{\alpha},H_{q}\rangle=\vec{\alpha}|_{q}\;.
\end{equation}
The rank of $\mathfrak{su}(N)$ is $r=N-1$, and its dimension is
$d=N^{2}-1$. The weights of the defining representation, can be ordered
according to, 
\begin{equation}
\vec{w}_{1}>\vec{w}_{2}>\dots>\vec{w}_{N}\;,
\end{equation}
so that the positive and simple roots are, respectively, 
\begin{equation}
\vec{\alpha}_{qp}=\vec{w}_{q}-\vec{w}_{p}\makebox[.3in]{,} q<p
\makebox[.5in]{,} \vec{\alpha}_{q}=\vec{w}_{q}-\vec{w}_{q+1}
\label{difw}\;.
\end{equation} 
Finally, recalling that the fundamental weights $\vec{\Lambda}^{q}$ are defined
by, 
\begin{equation}
\frac{2\,\vec{\alpha}_{q}\cdot\vec{\Lambda}^{p}}{\vec{\alpha}_{(q)}\cdot\vec{\alpha}_{(q)}}=\delta_{q}^{\, p}\;,
\end{equation}
the $\vec{\lambda}^{q}$ basis in eq. (\ref{alam}) can be written 
as, 
\begin{equation}
\vec{\lambda}^{q}=2N\,\vec{\Lambda}^{q}\;,
\end{equation}
with, 
\begin{equation}
\vec{\Lambda}^{1}=\vec{w}_{1}\makebox[.3in]{,}\vec{\Lambda}^{2}=\vec{w}_{1}+\vec{w}_{2}\makebox[.3in]{,}\vec{\Lambda}^{3}=\vec{w}_{1}+\vec{w}_{2}+\vec{w}_{3}\makebox[.3in]{,}\dots
\end{equation}

\bibliographystyle{h-physrev4}

\end{document}